# The cosmic ray muon tomography facility based on large scale MRPC detectors

WANG Xuewu [a,b], ZENG Ming [a,b,*], ZENG Zhi [a,b], WANG Yi [a,b], ZHAO Ziran [a,b], YUE Xiaoguang [a], LUO Zhifei [a],

YI Hengguan [a], YU Baihui [a], CHENG Jianping [b]

[a] *Department of Engineering Physics, Tsinghua University, Beijing, 100084, China*

[b] *Key Laboratory of Particle & Radiation Imaging (Tsinghua University), Ministry of Education, Beijing, 100084, China*

**Abstract:** Cosmic ray muon tomography is a novel technology to detect high-Z material. A prototype of TUMUTY with 73.6 cm x 73.6 cm large scale position sensitive MRPC detectors has been developed and is introduced in this paper. Three test kits have been tested and image is reconstructed using MAP algorithm. The reconstruction results show that the prototype is working well and the objects with complex structure and small size (20 mm) can be imaged on it, while the high-Z material is distinguishable from the low-Z one. This prototype provides a good platform for our further studies of the physical characteristics and the performances of cosmic ray muon tomography.

## 1. Introduction

Cosmic ray muon tomography is a novel technology for high-Z material detection with high penetration and intrinsic safety feature [1]. Several muon imaging facilities had been setup all over the world based on different types of detectors, such as drift tube, drift chamber, GEM, RPC, scintillator and so on[2-5]. Since 2011, cosmic ray muon tomography group at Tsinghua University have been developing the prototype of Tsinghua University cosmic ray MUon Tomography facilitY (TUMUTY). 73.6 cm x 73.6 cm large scale 2D position sensitive MRPC detectors are adopted because of their advantages: large area, high time resolution and low cost. Especially, a fine-fine encoding readout electronics had been developed to deal with the 2688 channels of induced signals, which reduced the complexity of the system significantly [2]. The first image of TUMUTY had been reconstructed in 2013. In this paper, three test kits have been tested and are reconstructed with MAP algorithm to evaluate the performance of the

---

complex structure reconstruction, the spatial resolution and the material discrimination of TUMUTY.

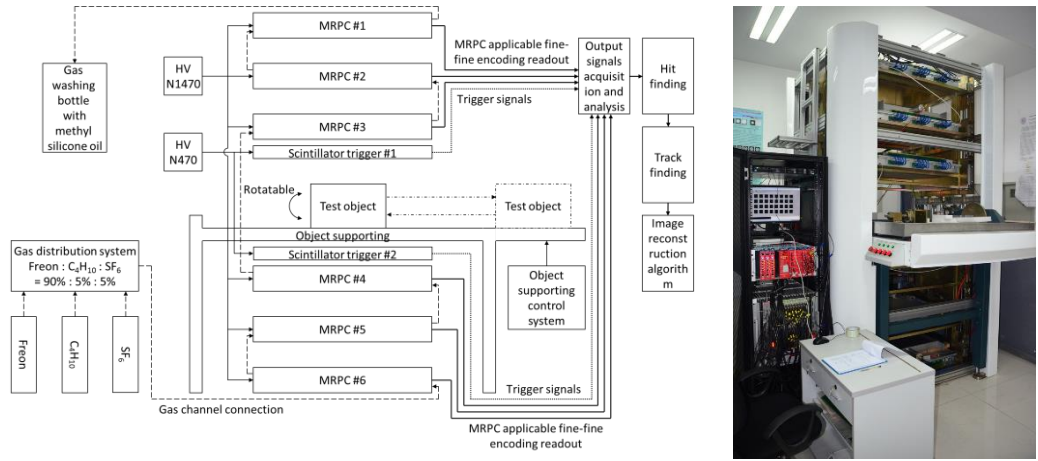

Fig. 1 the system construction (left) and the photo (right) of TUMUTY

## 2. The prototype of TUMUTY

### 2.1. MRPC Detectors

As shown in Fig. 1, there are 6 layers of large scale 2D position sensitive MRPC detectors in TUMUTY, with 224 channels of induced signals on each dimension, 2688 channels for the whole facility. The structure of the MRPC detector is briefly illustrated in Fig. 2, more details can be found in Ref[2]. A gas mixture which contains 90% freon gas R134A, 5% iso-butane and 5% sulfur hexafluoride is used as the working gas. From the inside out, it is composed of 5 slices of the inner glass with 0.55 mm thickness, 2 slices of the outer glass with 0.7 mm thickness, carbon films, Mylar films, PCB plates and honeycomb plates. The 6 layers of the gas gaps divided by 5 slices of the inner glass are 0.25 mm thick each. The carbon films adhered to the outer glass is used as high voltage electrodes while the Mylar films act as insulators between the high voltage electrodes and the PCB plates. The charge-induced signals are acquired from the Cu readout strips attached to the PCB plates. The honeycomb plates play the roles of supporting and protection. After packaged into aluminum gas-box, the detectors are assembled into a steel bracket which provides mechanical support and adjustable spacing between detectors. An independent mechanical holder to insert the test kits into the center of the detectors and rotate the kits is also established in TUMUTY.

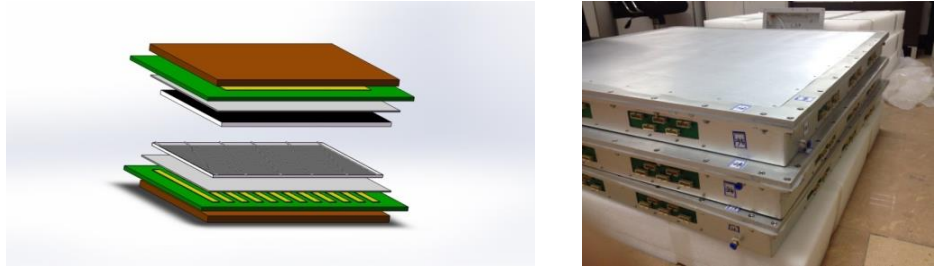

Fig. 2 the structure of the MRPC detector

## 2.2. Readout Electronics

The 8-ch current sensitive fast preamp ASIC and high speed waveform sampling combined with digital pulse process technology is adopted. To reduce the scale of readout electronics, a fine-fine encoding readout method has been developed, and 30 channels (16+14) readout electronics were used to deal with the 224 channels induced signals of each dimension. The method was originally used in the MCP detector as the MAMA (Multi-Anode Microchannel Array) which was used in the Hubble Space Telescope [6]. A mathematical model of the fine-fine encoding readout method has been developed and applied to find out the best design. The schematics of the readout electronics are illustrated in the left image in Fig. 3. Triggers are provided by the coincidence output of a pair of scintillators placed at the top and bottom of TUMUTY. When a trigger signal is received, the output of shaping circuit will be digitized by the DAQ module and all the data obtained from detectors will be transferred and saved to a computer via Ethernet. And digitized pulse shapes of 30 channels readout electronics for an actual muon event is showed in the right image in Fig. 3.

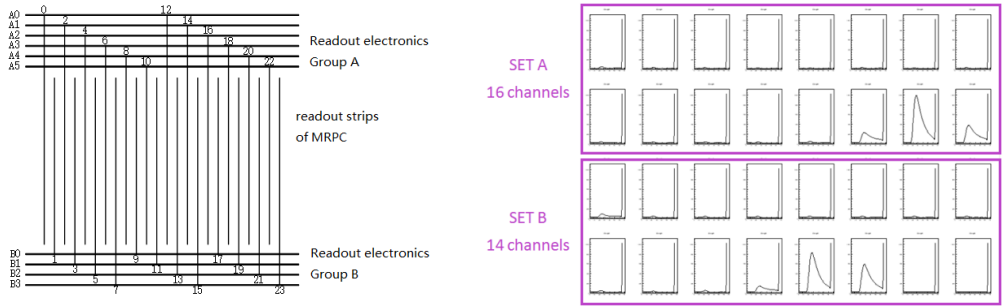

Fig. 3 a fine-fine encoding readout electronics (left) and an actual muon event (right)

## 3. Data analysis and results

### 3.1. Hit reconstruction

The event data are analyzed offline to determine the muon hit position in each MRPC detector. With several steps including converting original waveform data to the ROOT tree format [8], removing the pedestal value, charge integral and fine-fine decoding, the original event position can be reconstructed from the 30 readout signal pulse shapes with a decoding lookup table and charge centroid method. The hit positions on each layer of detector then can be used to generate the 3D tracks of the muons.

### 3.2. Track finding

The position offsets between the MRPC detectors need to be corrected before further process. As the position offsets are static, they can be measured with cosmic muon data without test kits.

After detector position offsets are corrected, 3D track of muons can be precisely reconstructed with principal component analysis [9] and least square method.

### 3.3. Test kits

In the previous work, several models including the Tsinghua University Cube for Cargo Inspection

(TUCCI) model has been proposed and studied with Monte Carlo simulation [10]. In this work, 3 test kits were imaged in a single run on TUMUTY to evaluate the performance of the complex structure reconstruction, the spatial resolution and the material discrimination, as shown in

. The supporting holder is made of polyethylene:

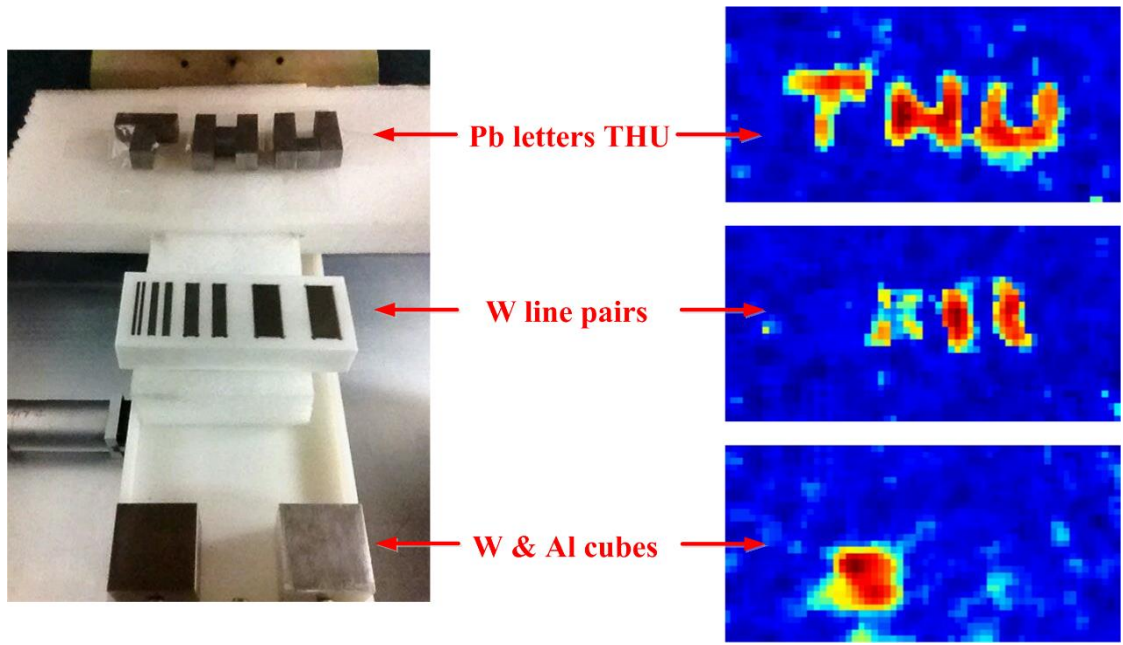

Fig. 4 the photograph of the test kits (left) and the MAP reconstruction result (right).

Test kit 1: the letters "THU" model (upper) consists of several lead bricks, and the size of each letter is about $60^2$ mm$^2$ with 20 mm thickness;

Test kit 2: the line pairs model (middle) is made of tungsten cubes, with widths and thicknesses of 2mm, 2mm, 5mm, 5mm, 10mm, 10mm, 20mm and 20mm. And the length is 50mm.

Test kit 3: two $50^3$ mm$^3$ cubes (bottom) are made of tungsten (left) and aluminum (right) respectively.

### 3.4. Image reconstruction

Nearly 900,000 events are collected by TUMUTY within 12 days, and the tracks of muons are fitted via PCA (Principal Component Analysis) method before reconstruction. The profile plot in the middle layer of the 3D reconstruction results reconstruction results are shown in Fig. 4. The reconstruction algorithm is MAP (Maximum a Posteriori) and the voxel size is set to 5mm. The outline of the "THU" letters are reconstructed clearly; the 10 mm line pairs can be distinguished but the edge is not well reconstructed; the cube made of tungsten can be identified while the aluminum cube is invisible due to its small density. The cross-sections in the middle layer of the 3D reconstruction results are presented. The results show that the objects with complex structure and the 20 mm line pairs can be reconstructed by TUMUTY system in 12 days, and the high atomic number material can be distinguished with the low one.

## 4. Conclusions

The cosmic ray muon tomography facility based on large scale MRPC detectors has been developed. Several models have been tested to evaluate the performance of this prototype. The reconstruction images demonstrate that the prototype of TUMUTY is working well and the objects with complex structure and small size (20 mm) can be reconstructed on it. The high-Z material is also distinguishable from the low-Z one. This prototype provides a good platform for our further studies of the physical characteristics and the performances of cosmic ray muon tomography.

## 5. Acknowledgments

This work is supported by National Natural Science Foundation of China (No.11035002 and No.11175099).